\documentclass[pra,aps,twocolumn,a4paper,floatfix,nofootinbib,notitlepage]{revtex4-1}

\usepackage[utf8]{inputenc}
\usepackage{graphicx,psfrag}
\usepackage{mathrsfs}
\usepackage{amsmath,amsfonts,amssymb}
\usepackage{comment}
\usepackage{xcolor}
\usepackage{enumerate}
\usepackage{array}
\usepackage{hyperref}
\usepackage{boldline,multirow}
\usepackage{lineno}
\usepackage{siunitx}
\usepackage{framed}
\hypersetup{
    colorlinks = true,
    linkcolor = {blue},
    citecolor = {blue},
    urlcolor = {blue},
    linkbordercolor = {white},
    }
\begin{document}

\title{Modeling of dual frequency combs and bistable solitons in third-harmonic generation}

\author{Tobias \surname{Hansson}$^{1,2,*}$}
\author{Pedro \surname{Parra-Rivas}$^2$} 
\author{Stefan \surname{Wabnitz}$^2$}

\affiliation{${}^1$ Department of Physics, Chemistry and Biology,
Link\"oping University, SE-581 83 Link\"oping, Sweden}
\affiliation{${}^2$ Dipartimento di Ingegneria dell'Informazione, Elettronica e Telecomunicazioni, Sapienza Universit\`a di Roma, via Eudossiana 18, 00184 Rome, Italy}
\affiliation{${}^*$ Corresponding author: tobias.hansson@liu.se}

\date{\today}

\begin{abstract}
\vspace{.1cm}
\section*{Abstract}
Phase-matching of the third-harmonic generation process can be used to extend the emission of radiation from Kerr microresonators into new spectral regions far from the pump wavelength. Here, we present a theoretical mean-field model for optical frequency combs in a dissipative and nonlinear $\chi^{(3)}$-based cavity system with parametric coupling between fundamental and third-harmonic waves. We investigate temporally dispersive dual-comb generation of phase-matched combs with broad bandwidth, and report conditions for accessing a multistable regime that simultaneously supports two types of coupled bright cavity solitons. These bistable cavity solitons coexist for the same pump power and frequency detuning, while featuring dissimilar amplitudes of their individual field components. Third-harmonic generation frequency combs can permit telecom pump laser sources to simultaneously directly access both the near-infrared and the visible regions, which may be advantageous for the development of optical clocks and sensing applications. 
\end{abstract}

\maketitle

\section*{Introduction}

Optical frequency combs (OFCs) utilize the nonlinear polarization response of a cavity-enclosed dielectric medium, in order to convert an externally applied pump field to multiple new frequencies. Acting as broadband and coherent optical sources, OFCs are a key technology for enabling a diverse range of applications such as frequency metrology, optical communications and spectroscopy \cite{Udem2002,Kippenberg2011,Pasquazi2018}. However, conventional Kerr comb synthesizers only emit radiation in a spectral range that is centered around the pump laser frequency, and generally require anomalous group-velocity dispersion for the experimentally accessible formation of phase-locked states \cite{Agha2007,Matsko2012}. This makes it challenging to form combs in wavelength ranges which lack suitable pump laser sources, and in spectral regions that exhibit an effective waveguide and material dependent normal dispersion.

One way of overcoming these limitations is to exploit the third-harmonic generation (THG) process of the $\chi^{(3)}$-nonlinearity, in order to couple pump field excitations to parametric waves at thrice the fundamental frequency (FF), $3\omega_1$. The THG process is inherent in all transparent nonlinear media that display a strong Kerr effect, but in practice it is hampered by the requirement of maintaining a fixed phase relation, which is necessary for efficient frequency conversion \cite{Boyd}. While many experimental observations of THG in microcomb devices have relied on refractive index matching between the fundamental and higher-order whispering gallery modes, it is also feasible to accomplish phase-matching through birefringence, periodic poling and other quasi-phase-matching techniques \cite{Helmy2011}.

In this work we consider a centrosymmetric nonlinear Kerr resonator system that is engineered to phase-match the third-harmonic process in order to enable resonant dual-comb generation around both the FF and the third-harmonic (TH) frequency, when the dissipative cavity is driven by a continuous-wave (CW) pump source at the fundamental frequency $\omega_1$. We go beyond previous studies of cavity THG that have been restricted to the non-dispersive case with only two interacting frequencies \cite{Rodriguez2007,Li2018}, by considering the mutual coupling between sidebands around each carrier wave and the simultaneous interaction of all frequency modes. This system shares similarities with non-phase-matched Kerr microresonators, that can be modelled by the Lugiato-Lefever equation (LLE) \cite{Lugiato1987,Coen2013} or driven-and-damped nonlinear Schr\"odinger equation \cite{Hael92}; it is also analogous to OFCs in quadratically nonlinear resonators, which exploit cascaded processes, found in $\chi^{(2)}$-nonlinear media without inversion symmetry, in order to enable coupling and dual-comb generation around both the FF and the second-harmonic frequency \cite{Ricciardi2020,Xue2017}. We note that a similar model of THG-assisted four-wave mixing was published in Ref.~\cite{Zhang2022} during the final preparation of this manuscript, but with the inclusion of simplifying assumptions that limit its applicability to a perturbative THG regime, and exclude the possibility of generating bistable cavity solitons. 

Previous experimental work has demonstrated the direct emission of visible light by THG from an infrared pump using both high-Q whispering-gallery-mode and integrated microresonators \cite{Carmon2007,Farnesi2014,Wang2016,Surya2018,Pampel2020}. The generation of OFCs by THG acting together with Raman-assisted spectral broadening in silica based microcavities was also reported \cite{Tanabe2016}. Theoretical studies of spatially diffractive beam propagation in conservative, cavityless systems have also shown the possibility of generating both bright and dark coupled solitary wave structures in the presence of THG; that exhibit properties such as a power threshold and bistability \cite{Sammut1997,Sammut1998}. Given that the dynamics of our system is governed solely by the fundamental field in the limit of vanishing parametric coupling, one may expect to find a similar homotopic extension of the phase-locked bright cavity solitons (CSs) that are supported by the LLE in the case of anomalous group-velocity dispersion \cite{Leo2010,Herr2013}. Surprisingly, we find that coupled FF and TH combs can also support an additional type of CS with a partially overlapping range of existence.
These dual, two component, solitons share a common trapping refractive index potential through self- and cross-phase modulation (SPM/XPM). Moreover, there is the intriguing possibility that mutual XPM coupling can be used to overcome the group-velocity mismatch, in order to create various types of synchronized dual-frequency combs with locked repetition rates around both the FF and the TH frequency.

In the following sections, we develop a theoretical mean-field model for a doubly-resonant, cavity-enhanced, dispersive and nonlinear system, phase-matched for THG. In particular we show that, as the TH field grows larger, it may not be simply considered as an up-converted replica of the fundamental comb, but it will reciprocally influence the latter through both parametric coupling and XPM. We investigate the multistability properties of the homogeneous solution and consider the importance of modulational instability (MI) in generating various types of multi-frequency combs. Additionally, we make a detailed numerical study of the multistable regime, where we demonstrate the occurrence of bistable cavity solitons that coexist, when both the FF and the TH frequency lie in the anomalous dispersion regime.

\section*{Results and Discussion}
\subsection*{Model}

\begin{figure}[!t]
    \includegraphics[width=\linewidth]{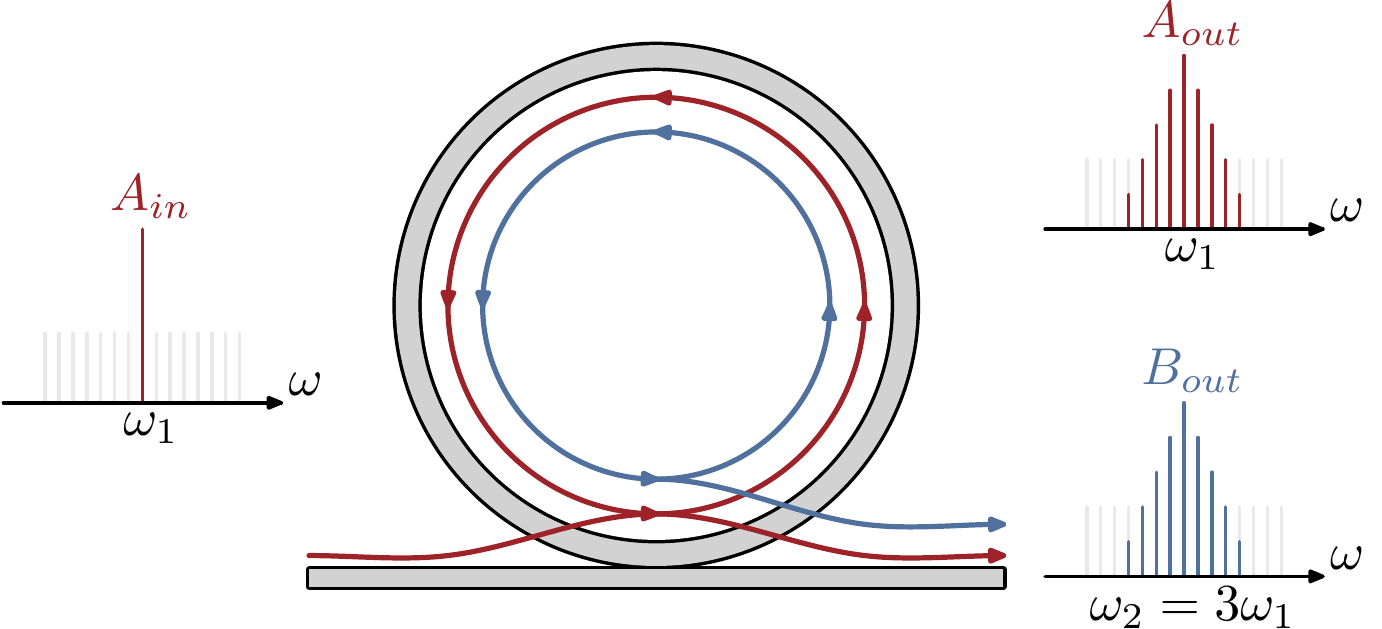}
    \caption{\textbf{Schematic of the THG resonator system.} The $\chi^{(3)}$-based nonlinear microresonator is phase-matched for third-harmonic generation. The resonator is driven by a CW pump field at the fundamental frequency $\omega_1$ and generates simultaneous frequency combs around both $\omega_1$ and $\omega_2 = 3\omega_1$.}
    \label{fig:chi3res}
\end{figure}

We consider OFC generation in a dispersive $\chi^{(3)}$-based resonator system with coupling between fundamental and third-harmonic fields, as schematically illustrated in Fig.~\ref{fig:chi3res}. The system is assumed to be resonant around both the driving frequency of the fundamental field (FF) $\omega_1$ as well as the frequency of the third-harmonic (TH) field $\omega_2 = 3\omega_1$. For simplicity, we assume an isotropic nonlinear polarization response $\bar{P}_{NL} = \epsilon_0\chi^{(3)}\bar{E}^3$ and a linearly polarized electric field propagating along the $z$-axis, viz.
\begin{align}
   \bar{E} = &\hat{e}\frac{1}{2}\Big(F_1(x,y)A(z,t)e^{i(k_1z-\omega_1t)} + \nonumber\\ 
   & F_2(x,y)B(z,t)e^{i(k_2z-\omega_2t)}\Big) + c.c.
\end{align}
where $\epsilon_0$ is the vacuum permittivity, $\chi^{(3)}$ is the third-order nonlinear susceptibility, $\hat{e}$ is a unit vector in the polarization direction and $c.c.$ denotes complex conjugate. The OFC generation dynamics is modelled by means of a scalar Ikeda map \cite{Ikeda1979} for the evolution of the temporal field during each roundtrip, together with appropriate boundary conditions for the injection of the external pump and the coupling of the fields from one roundtrip to the next. Starting from Maxwell's equations and applying the slowly-varying envelope approximation, the envelopes of the co-propagating electric field of the fundamental $A_m$ and the third-harmonic $B_m$ at the $m$th roundtrip are found to obey the following coupled nonlinear equations:
\begin{align}
  & \frac{\partial A_m}{\partial z} = \left[-\frac{\alpha_{c1}}{2} + iD_1\left(i\frac{\partial}{\partial t}\right)\right]A_m + i\frac{\omega_1n_2(\omega_1)}{c}\times \nonumber\\ 
  & \left[Q_{13}B_m(A_m^*)^2e^{-i\Delta k z} + \left(Q_{11}|A_m|^2 + 2Q_{12}|B_m|^2\right)A_m\right], \label{eq:cNLS1} \\
  & \frac{\partial B_m}{\partial z} = \left[-\frac{\alpha_{c2}}{2} + iD_2\left(i\frac{\partial}{\partial t}\right)\right]B_m + i\frac{\omega_2n_2(\omega_2)}{c}\times \nonumber\\ 
  & \left[\frac{Q_{23}}{3}A_m^3e^{i\Delta k z} + \left(2Q_{21}|A_m|^2 + Q_{22}|B_m|^2\right)B_m\right], \label{eq:cNLS2}
\end{align}
where $z$ is the longitudinal coordinate and $t$ is time. The dispersive properties of the medium that are associated with the non-equidistant resonance spacing are described by the Taylor series expansions $D_{1,2}(i\partial/\partial t) = \sum_{n=1}^\infty(k_{1,2}^{(n)}/n!)(i\partial/\partial t)^n$ of the propagation constants $k_{1,2}(\omega)$ with $k_{1,2}^{(n)} = d^nk_{1,2}/d\omega^n|_{\omega_1,\omega_2}$.
Here $k_{1,2}'$ are inverse group-velocities, $k_{1,2}''$ are group-velocity dispersion coefficients and $\Delta k = 3k_1 - k_2$ is a wave-vector mismatch. Moreover, we have that $Q_{ij}$ are modal overlap integrals, $\alpha_{c1,2}$ are the FF/TH absorption losses, $c$ is the speed of light in vacuum and $n_2(\omega) = 3\chi^{(3)}(\omega)/8n(\omega)$ is the nonlinear coefficient, with $n(\omega)$ the linear refractive index. In the case of natural phase-matching we have $\Delta k = 0$ which requires matching of the FF/TH refractive indices $n(\omega_1) = n(\omega_2)$ and implies that $\Delta_2 = 3\Delta_1$, which is assumed in the following.
The transverse overlap integrals that are needed to account for the variation in spatial mode profiles between families of different mode orders are given by
\begin{align}
  & Q_{11} = \frac{\int|F_1|^4dS}{(\int|F_1|^2dS)^2}, \qquad Q_{12} = \frac{\int|F_1|^2|F_2|^2dS}{(\int|F_1|^2dS)(\int|F_2|^2dS)}, \nonumber\\
  & Q_{13} = \frac{\int(F_1^*)^3F_2dS}{(\int|F_1|^2dS)^{3/2}(\int|F_2|^2dS)^{1/2}}, \nonumber\\
  & Q_{21} = \frac{\int|F_1|^2|F_2|^2dS}{(\int|F_1|^2dS)(\int|F_2|^2dS)}, \qquad Q_{22} = \frac{\int|F_2|^4dS}{(\int|F_2|^2dS)^2}, \nonumber\\
  & Q_{23} = \frac{\int F_1^3F_2^*dS}{(\int|F_1|^2dS)^{3/2}(\int|F_2|^2dS)^{1/2}},
\end{align}
where $Q_{21} = Q_{12}$, $Q_{23}^* = Q_{13}$ and $dS = dxdy$. These definitions reduce to the familiar Kerr coefficient $\gamma = \omega_1n_2(\omega_1)/cA_{eff}$ with $A_{eff} = Q_{11}^{-1}$ in the absence of any TH field.

The fields at the beginning of the $(m+1)$th roundtrip are assumed to be related to the fields at the end of the $m$th roundtrip through the boundary conditions
\begin{align}
  & A_{m+1}(0,t) = \sqrt{\theta_1}A_{in} + \sqrt{1-\theta_1}e^{-i\delta_1}A_m(L,t), \label{eq:BC1}\\
  & B_{m+1}(0,t) = \sqrt{1-\theta_2}e^{-i\delta_2}B_m(L,t), \label{eq:BC2} 
\end{align}
that model a generic optical coupling, such as the evanescent field overlap from a nearby waveguide or tapered fiber, that partially transmits the external pump field $A_{in}$ while reflecting the intracavity fields from the previous roundtrip. Here, $L$ is the length of the cavity, while $\theta_{1,2}$ are the power transmission coefficients and $\delta_{1,2}$ are the phase detunings of the FF/TH fields from the nearest cavity resonance. We note that the complementary case of a down-converting, $3\omega_1\to\omega_1$, optical parametric oscillator can be modelled by simply moving the pump term to Eq.~(\ref{eq:BC2}) for the TH field.

The Ikeda map constitutes a complete model for the temporal and spectral dynamics of a THG cavity based OFC generation system for general resonance and phase-matching conditions. But analytical and numerical investigations can be simplified in the doubly-resonant case ($\theta_{1,2} \approx 1$) by averaging the above map over the roundtrip length into a mean-field model, similar to the LLE. In the following, we truncate the dispersion to the second-order; assume the phase-matching to be almost perfect, so that the coherence length is longer than the cavity length; and shift to a retarded reference frame moving with the group-velocity of the driving field $(k_1')^{-1}$. Following a derivation, whose details are presented in the Methods section, we obtain our main system of normalized mean-field evolution equations for the FF and TH fields $A$ and $B$ as
\begin{align}
  \frac{\partial A}{\partial t} = & \left[-(1+i\Delta_1) - i\eta_1\frac{\partial^2}{\partial\tau^2}\right]A + \nonumber\\
  & i\left[\kappa^*B(A^*)^2 + (|A|^2 + 2\sigma|B|^2)A\right] + f, \label{eq:A}\\
  \frac{\partial B}{\partial t} = & \left[-(\alpha+i\Delta_2) - d\frac{\partial}{\partial\tau}  - i\eta_2\frac{\partial^2}{\partial\tau^2}\right]B + \nonumber\\
  & i3\rho\left[\frac{\kappa}{3}A^3 + (2\sigma|A|^2 + \mu |B|^2)B\right], \label{eq:B}
\end{align}
where $t$ and $\tau$ are normalized slow- and fast-time variables, respectively. $\alpha = \alpha_2/\alpha_1$ is the ratio of the roundtrip losses, $\Delta_j = \delta_j/\alpha_1$ is the normalized detuning, $d = \sqrt{2L/(|k_1''|\alpha_1)}\Delta k'$ is the walk-off parameter that depends on the group-velocity mismatch $\Delta k' = k_2'-k_1'$, $\eta_j = k_j''/|k_1''|$ is the ratio of the group-velocity dispersion coefficients and $f = \sqrt{\theta_1\omega_1n_2(\omega_1)LQ_{11}/(c\alpha_1^3)}A_{in}$ is the normalized input pump field. The nonlinear interaction is governed by the four dimensionless parameters
\begin{align}
  & \rho = \frac{n_2(\omega_2)}{n_2(\omega_1)}, \qquad \mu = \frac{Q_{22}}{Q_{11}}, \qquad \sigma = \frac{Q_{21}}{Q_{11}}, \nonumber\\
  & \kappa = \frac{Q_{23}}{Q_{11}}e^{i\Delta kL/2}\textrm{sinc}(\Delta kL/2),
\end{align}
that can be assumed to be close to unity, unless the phase-matching is significantly multimodal. It is interesting to note that Eqs.~(\ref{eq:A}-\ref{eq:B}) are formally similar to models describing phase-matched doubly-resonant second-harmonic generation (SHG) in quadratic nonlinear media with a simultaneous Kerr nonlinearity \cite{Xue2017,Villois2019}. The two systems differ mainly in the magnitude of the terms and in the exponents of the FF that appears in the parametric coupling terms: these read as $BA^*$ and $A^2$ in the case of second-harmonic generation.

\subsection*{Homogeneous solutions}

\begin{figure}[!t]
    \includegraphics[width=\linewidth,trim={0.5cm 1.0cm 1.5cm 2.0cm},clip]{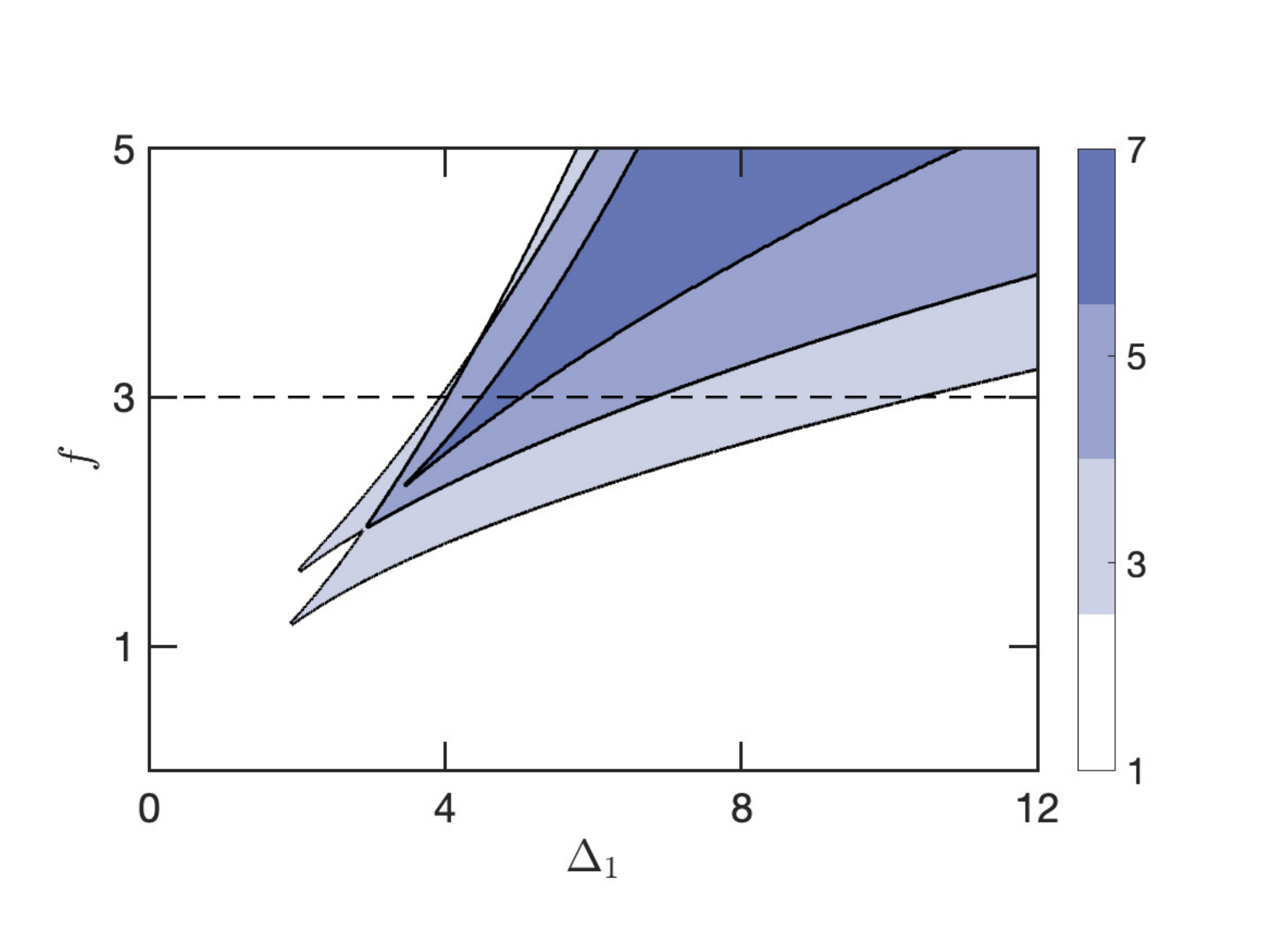}
    \caption{\textbf{Multistability of homogeneous solutions.} The plot shows colored parameter regions with 1,3,5 or 7 simultaneous homogeneous solutions that coexist for the same normalized detuning $\Delta_1$ and pump power $f$.}
    \label{fig:bistability}
\end{figure}
\begin{figure*}[!t]
  \begin{minipage}{\linewidth}
    \centering 
    \includegraphics[width=\linewidth]{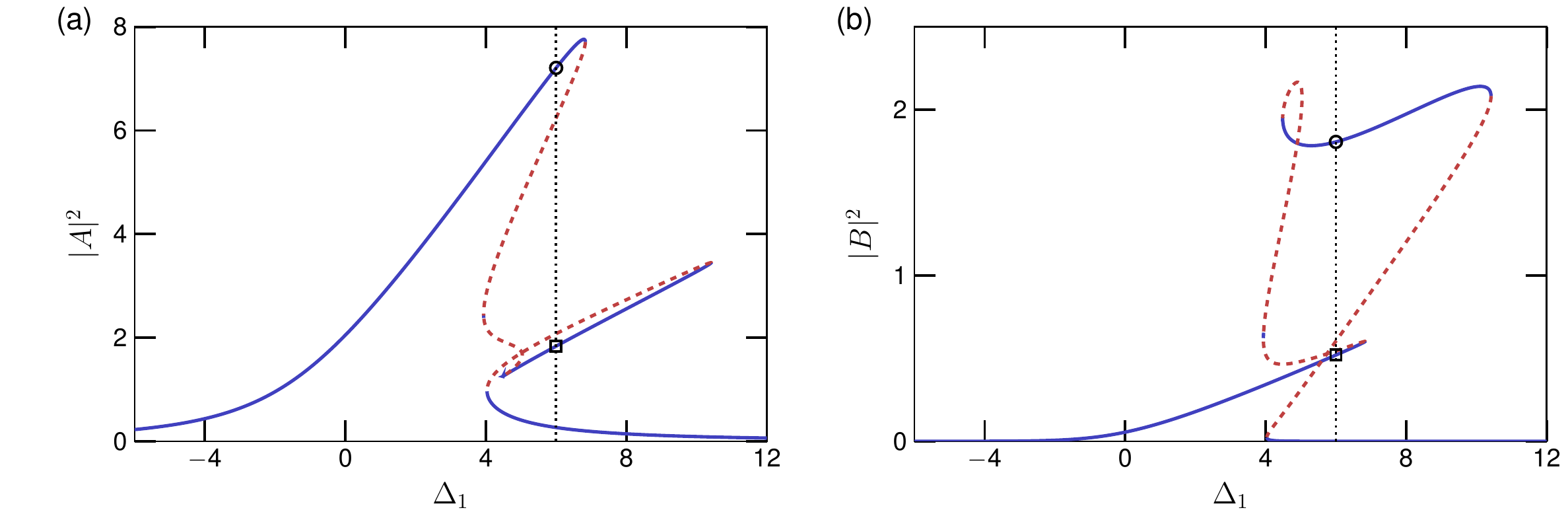}
    \caption{\textbf{Multistable resonance response.} The plots show intracavity power of the homogeneous solution as a function of detuning when the cavity is driven by an external pump with normalized power $f = 3$. Panels (a) and (b) show the intracavity power of the FF and TH fields, respectively, where blue and red colors denote branches that are stable/unstable to homogeneous perturbations. The dotted vertical line at $\Delta_1 = 6$ indicates the location of the bistable cavity solitons considered below.}
    \label{fig:res1}
  \end{minipage}
\end{figure*}
The response of the system for pump powers below the threshold for parametric comb generation is characterized by CW emission at both FF and TH frequencies. We find a set of stationary mixed-mode homogeneous solutions by setting the derivatives in Eqs.~(\ref{eq:A}-\ref{eq:B}) to zero. Eliminating terms that are linear in $B$ one finds that
\begin{align}
  & \left[(1+a_r)+i(\Delta_1-a_i)\right]A = f, \label{eq:res1}\\
  & \left[\alpha + i(\Delta_2-b)\right]B = i\rho\kappa A^3, \label{eq:res2}
\end{align}
where we have defined the power dependent functions
\begin{align}
  & a_r = \frac{\alpha}{\rho}\frac{|B|^2}{|A|^2}, \qquad a_i = \frac{a_r}{\alpha}\left(\Delta_2-b\right) + \left(|A|^2+2\sigma|B|^2\right), \nonumber\\
  & b = 3\rho\left(2\sigma|A|^2+\mu|B|^2\right).
\end{align}

Eqs.~(\ref{eq:res1}-\ref{eq:res2}) are written in a resonance form, where a maximum occurs for a detuning that makes the imaginary part zero. The stationary fields are related through the power conservation law $f\left(A + A^*\right) = 2|A|^2 + 2(\alpha/\rho)|B|^2$, and the two FF/TH intracavity powers $|A|^2$ and $|B|^2$ are seen to satisfy a closed set of real nonlinear equations, viz.
\begin{align}
  & \left[(1+a_r)^2 + (\Delta_1-a_i)^2\right]|A|^2 = f^2, \label{eq:Ia}\\
  & \left[\alpha^2 + (\Delta_2-b)^2\right]|B|^2 = \rho^2|\kappa|^2|A|^6, \label{eq:Ib}
\end{align}
where as before $\Delta_2 = 3\Delta_1$ in the case of natural phase-matching. A detailed bistability analysis is complicated, owing to the high-order of the coupled Eqs.~(\ref{eq:Ia}-\ref{eq:Ib}). However, the equations can be solved numerically, in order to determine their number of solutions, as shown in Fig.~\ref{fig:bistability}. Here, we observe the presence of multistability, with an odd number of simultaneous solutions (1, 3, 5, or 7) in different ranges. We find no separate bistability of the TH: only a single TH solution corresponds to each value of the FF. In fact, because of pump depletion, as well as self- and cross-phase modulation, the solution for the TH field is not simply proportional to $A^3$ at high powers, but it can be expressed as an explicit function of the FF through the relation
\begin{align}
  & B = -\frac{\kappa A}{2\sigma}\frac{f-c_1A+ic_2|A|^2A}{f-c_3A^*+ic_4|A|^2A^*}, \nonumber\\
  & c_1 = 1+i\Delta_1, \qquad c_3 = 1-i\Delta_1+\frac{2\sigma}{3\rho\mu}(\alpha+i\Delta_2), \nonumber\\
  & c_2 = \frac{4\sigma^2}{3\mu}+1, \qquad c_4 = \frac{|\kappa|^2}{2\sigma}+\frac{4\sigma^2}{\mu}-1.
\end{align}

In Fig.~\ref{fig:res1} we show an example that illustrates the TH modification of the homogeneous solution as a function of the FF detuning $\Delta_1$ for $f=3$. As can be seen, the FF exhibits a Kerr tilted resonance shape, similar to that of the LLE model, but with a secondary peak associated with the resonance of the TH at higher values of the detuning. In particular, for a detuning around $\Delta_1 = 6$ we observe a multistable range with 5 different solutions, with three separate branches that are found to be stable to homogeneous perturbations (solid curves in Fig.~\ref{fig:res1}).

\subsection*{Modulational instability analysis}

The stability of the homogeneous solutions against periodic perturbations is important for determining the onset of comb generation and the accessibility of different solution branches. A positive MI gain causes the spontaneous growth of signal and idler sidebands that seed a cascade of phase-dependent four-wave mixing processes, eventually leading to broadband comb formation through the interplay of nonlinearity and dispersion. We analyze the MI gain by linearizing Eqs.~(\ref{eq:A}-\ref{eq:B}) around the homogeneous solution. Using the ansatz $A = A_0 + a_1e^{i\Omega\tau} + a_2e^{-i\Omega\tau}$ and $B = B_0 + b_1e^{i\Omega\tau} + b_2e^{-i\Omega\tau}$ we have the linear system $dw/dt = Mw$ with $w = [a_1, a_2^*, b_1, b_2^*]^T$ and the $4\times 4$ coefficient matrix
\begin{figure*}[!t]
  \begin{minipage}{\linewidth}
    \centering
    \includegraphics[width=\linewidth]{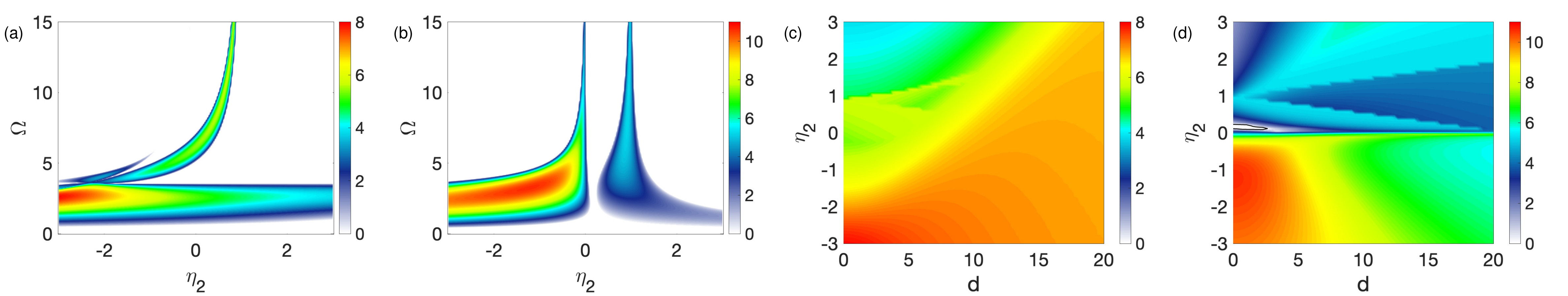}
    \caption{\textbf{Instability diagrams for the homogeneous solution.} Maximum MI growth rates for the multistable homogeneous solution in Fig.~\ref{fig:res1} as a function of TH dispersion $\eta_2$ and frequency $\Omega$, panels (a),(b), and walk-off $d$, panels (c),(d). The panels (a),(c) and (b),(d) show MI growth rates for the CW stable upper and middle branches, respectively, while the lowest branch is unconditionally stable (not shown). Case of $\Delta_1 = 6$ and anomalous dispersion for the FF, $\eta_1 = -1$.}
    \label{fig:MI}
  \end{minipage}
\end{figure*}
\begin{equation}
  M = \left[
    \begin{array}{cccc}
      -1+iq_1 & p_1 & p_2 & p_3 \\
      p_1^* & -1-iq_1 & p_3^* & p_2^* \\
      p_4 & p_5 & -\bar{\alpha}+iq_2 & p_6 \\
      p_5^* & p_4^* & p_6^* & -\bar{\alpha}-iq_2
    \end{array} \right],
    \label{eq:M_matrix}
\end{equation}
where $q_1 = \eta_1\Omega^2 - \Delta_1 + 2(|A_0|^2+\sigma|B_0|^2)$, $q_2 = \eta_2\Omega^2 - \Delta_2 + 6\rho(\sigma|A_0|^2+\mu|B_0|^2)$ and $\bar{\alpha} = \alpha + id\Omega$, and the off-diagonal coupling elements are given by $p_1 = i(A_0^2+2\kappa^*A_0^*B_0)$, $p_2 = i(2\sigma A_0B_0^*+\kappa^*(A_0^*)^2)$, $p_3 = i2\sigma A_0B_0$, $p_4 = -3\rho p_2^*$, $p_5 = 3\rho p_3$ and $p_6 = i3\rho\mu B_0^2$. The characteristic equation for the eigenvalues $\lambda$ is somewhat unwieldy, but it can be written in a factorized form that allows for an explicit solution in certain limiting cases, see Methods. It is clear that the MI is independent of the absolute phase through the invariance $A_0 \to A_0e^{i\theta_0}$ and $B_0 \to B_0e^{i3\theta_0}$. However, the MI does depend on the relative phase of the fields, because of their coherent coupling. 

The presence of TH coupling together with SPM/XPM also causes the appearance of new complex modulational instabilities, that have no counterpart in the LLE model, c.f.~Ref.~\cite{Zhang2022}. To illustrate this, we calculated the MI growth rates for the two homogeneously stable branches with the highest power in the multistable solution of Fig.~\ref{fig:res1}. Figure \ref{fig:MI} shows the dependence of the maximum MI growth rate, $\textrm{max}\{\textrm{Re}[\lambda(\Omega)]\}$, that is attainable for a finite perturbation frequency $\Omega$, versus the normalized group-velocity dispersion $\eta_2$ and group-velocity mismatch $d$ of the TH. It is seen that both branches generally are unstable to periodic perturbations when the FF dispersion is anomalous, except for the middle branch that has a small stability window for normal dispersion around $\eta_2 = 0.2$ and $d = 0$. We therefore fulfill the conditions for spontaneous sideband amplification that are a prerequisite for comb formation, possibly leading to separate phase-locked patterns and solitons that are associated with each branch. We note that the upper branch is dominated by the FF and only has a weak dependence on the TH, while the FF/TH fields are comparable on the middle branch, c.f. Fig.~\ref{fig:res1}. This is reflected in the stability properties and the sensitivity to the sign of the TH dispersion in the latter case.

\subsection*{Bistable solitons}

\begin{figure*}[!t]
  \begin{minipage}{\linewidth}
    \centering
    \includegraphics[width=\linewidth]{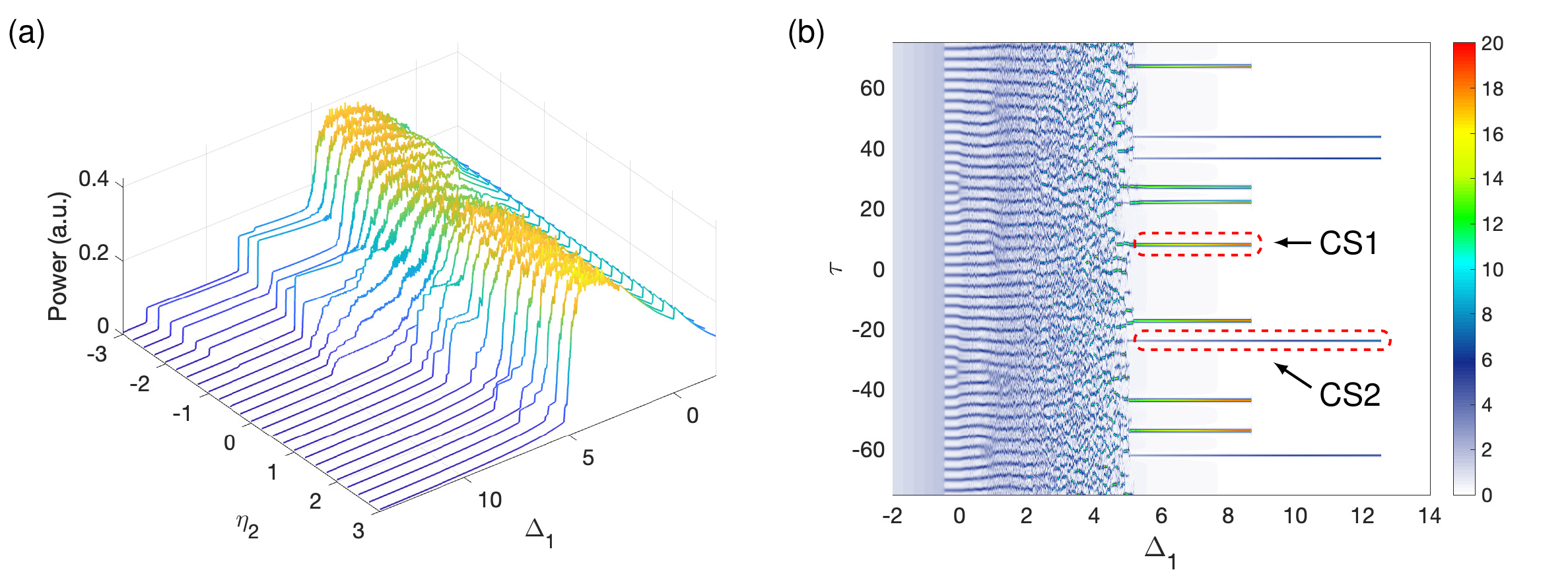}
    \caption{\textbf{Numerical simulation of detuning sweeps.} (a) Variation of comb power for different values of TH group-velocity dispersion $\eta_2$. The dispersion of the FF is assumed to be anomalous, $\eta_1 = -1$. (b) Temporal evolution of the FF power for the case of $\eta_2 = -2$. Characteristic signatures of a multisoliton regime with two different types of bistable cavity solitons indicated by CS1 and CS2 are seen for $\Delta_1 > 5$.}
    \label{fig:CS_sweep}
  \end{minipage}    
\end{figure*}
\begin{figure*}[!t]
\begin{minipage}{\linewidth}
    \centering
    \includegraphics[width=\linewidth]{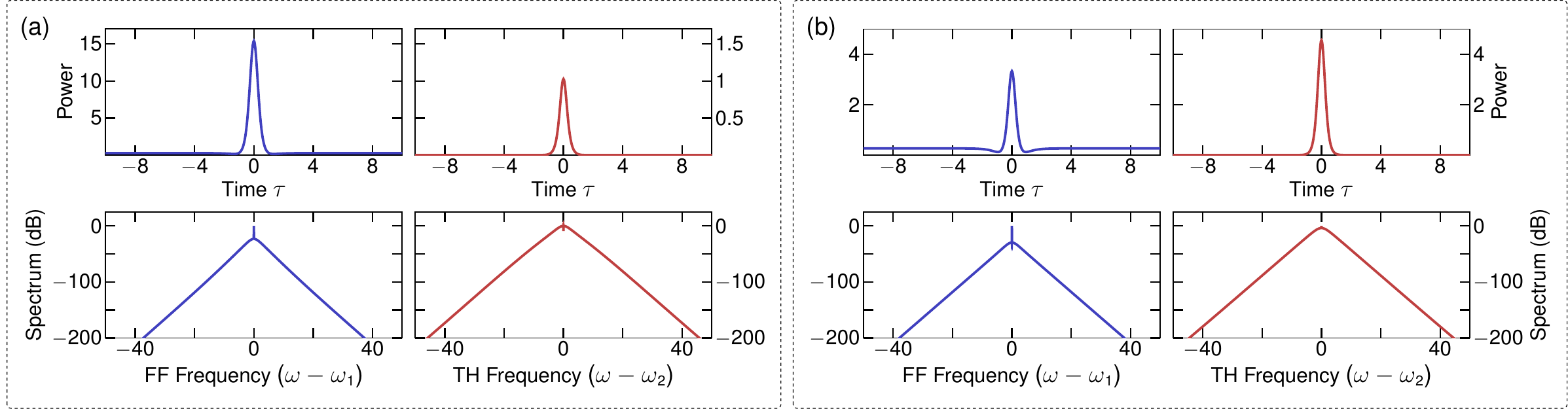}
    \caption{\textbf{Bistable soliton profiles.} Bistable solitons CS1 (a) and CS2 (b) existing for the same parameter values $f = 3$ and $\Delta_1 = 6$ with $\eta_1 = -1, \eta_2 = -2$ and $d = 0$. The top row shows temporal power profiles and the bottom row the corresponding frequency spectra. Blue and red colors denotes FF and TH components, respectively.}
    \label{fig:CS3}
  \end{minipage}
\end{figure*}
Next, let us investigate the fascinating possibility of finding bistable cavity soliton solutions in the multistable regime. Soliton bistability refers to the possibility of generating two (or more) localized structures with different temporal and spectral profiles, that coexist for the same values of pump power and cavity detuning. Such bistable (super) cavity solitons have previously been predicted in the framework of an Ikeda map for a Kerr medium without parametric coupling \cite{Hansson2015}, and have subsequently been experimentally confirmed to exist in fiber-ring resonators \cite{Anderson2017}. There it was established that bistable solitons form in regions of multistability, owing to the overlapping tilt of adjacent resonances at high pump powers. The coexistence of bistable vector solitons with different polarization states has also been found to occur in birefringent fiber-ring resonators \cite{Averlant2017,Kostet2021}. In general, we expect to find CSs in the vicinity of a detuning range where the homogeneous solution is bistable. Bright CSs are typically found in media with anomalous dispersion, and sit on a finite background that constitutes the lowest branch of the homogeneous solution. This background should be stable, while the upper branch may be modulationally unstable in favor of switching the stability to a periodic orbit corresponding to a stationary Turing pattern. A CS can then form by the locking of fronts that connect the homogeneous background and a cycle of the patterned state \cite{Parra_Rivas}.

We made a numerical search for CSs by performing a series of detuning sweeps over the resonance shown in Fig.~\ref{fig:res1}, for different values and signs of the TH dispersion parameter. We assumed $d = 0$, since walk-off leads to a separation of the pulse components, and is generally detrimental to soliton formation. The corresponding results are shown in Fig.~\ref{fig:CS_sweep}(a): here we plot the total comb power as a function of the FF cavity detuning $\Delta_1$ and the TH group-velocity dispersion $\eta_2$. We observe a dynamical sequence of evolving comb states with stable and chaotic MI regions followed by a series of steps that are characteristic of multi-soliton states \cite{Herr2013}. The length of the steps is found to vary with the sign and magnitude of the dispersion parameter, and is seen to be significantly longer in the case of anomalous dispersion ($\eta_2 < 0$). The termination point of the steps shows that only the upper branch of the homogeneous solution permits the formation of solitons for normal dispersion with $\eta_2 > 0$, while a sharp secondary step is observed at a detuning $\Delta_1 \approx 8.5$ for negative TH dispersion with $\eta_2 < -1$. This secondary step extends to a detuning $\Delta_1 \approx 12.5$, which lies beyond the endpoint of the middle branch in Fig.~\ref{fig:res1}.

In Fig.~\ref{fig:CS_sweep}(b) we show an example of the temporal evolution of the FF intracavity power for the case $\eta_2 = -2$, when both the FF and TH fields experience anomalous group-velocity dispersion. Here, we can identify a broad multi-soliton region that is split between the detuning ranges $\Delta_1 \approx 5-8.5$ and $\Delta_1 \approx 8.5-12.5$; and where in the first part we find traces of coexisting localized structures with two different amplitudes, i.e., bistable solitons. Isolated solitons corresponding to a detuning $\Delta_1 = 6$ are shown in Fig.~\ref{fig:CS3}. The two localized solitons may be associated with fronts that connect the homogeneous background with stationary states on the upper or middle branch, respectively. It is seen that the relative amplitudes of the FF/TH soliton components correspond to roughly twice the power of the homogeneous solutions. We observe that the power of the cavity soliton pair in panel (a) is dominated by the FF, with the TH being almost an order of magnitude smaller. Whereas the soliton pair in panel (b) has a much larger TH contribution: the total comb power is more equally split between the FF and TH components. It is notable that the background solution for the TH is very small in both cases, and has a spectral line magnitude that is comparable to that of the sidebands.
\begin{figure*}[!t]
  \centering
    \includegraphics[width=\linewidth]{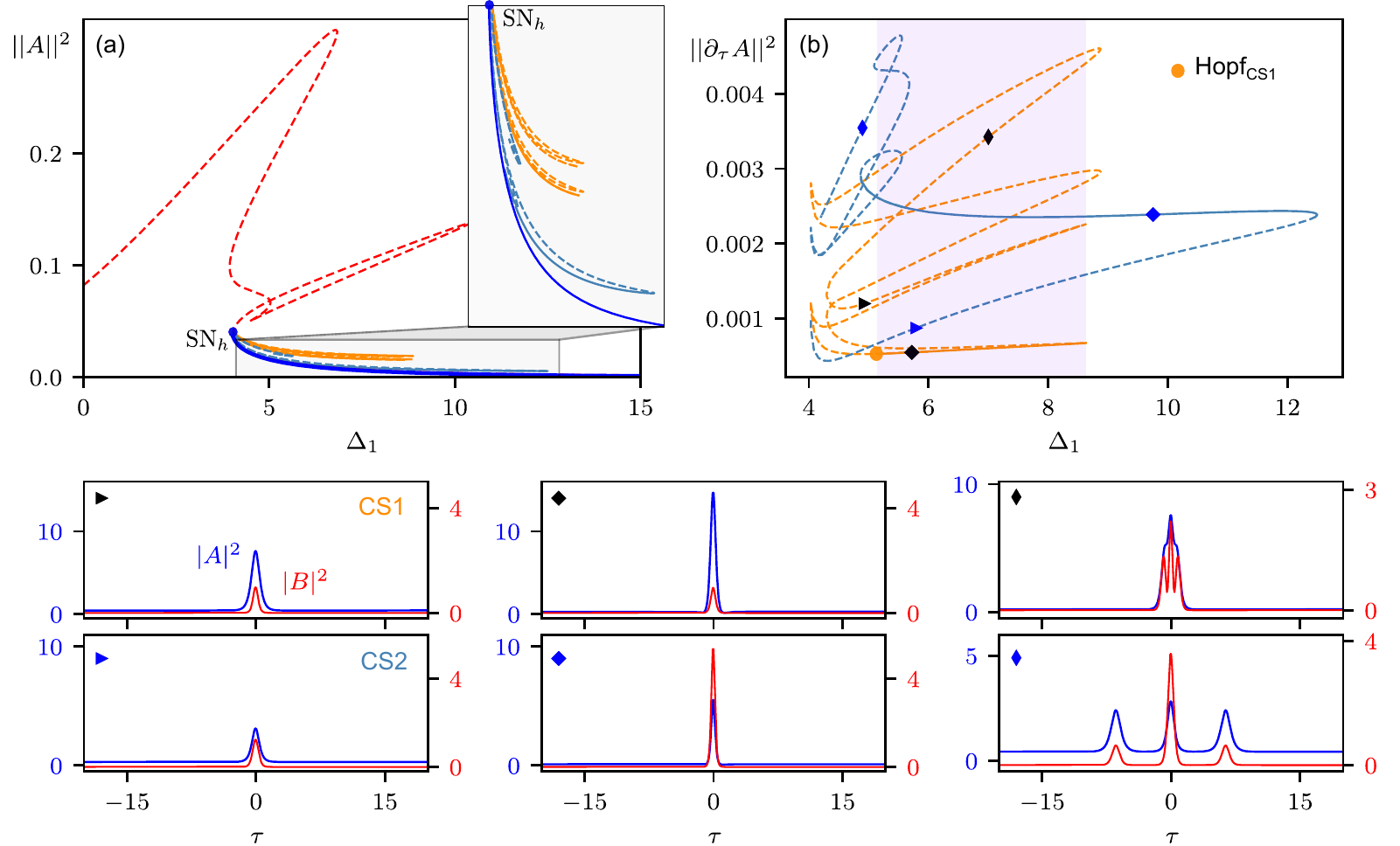}
  \caption{\textbf{Bifurcation structure of bistable CS solutions.} (a) Bifurcation diagram showing the modification of $||A||^2$ against $\Delta_1$ for $f=3$. (b) shows the same bifurcation curves as (a) but plotting $||\partial_\tau A||^2$ vs. $\Delta_1$. Orange and blue curves are associated with the solitons CS1 and CS2 of Fig.~\ref{fig:CS3}, respectively. In both panels, solid (dashed) lines correspond to stable (unstable) states. The bistability region between CS1 and CS2 is marked using a shadowed box. The different symbols correspond to the temporal profiles shown below.}
  \label{fig:CS_trace}
\end{figure*}
\begin{figure*}[!t]
\begin{minipage}{\linewidth}
    \centering
    \includegraphics[width=\linewidth]{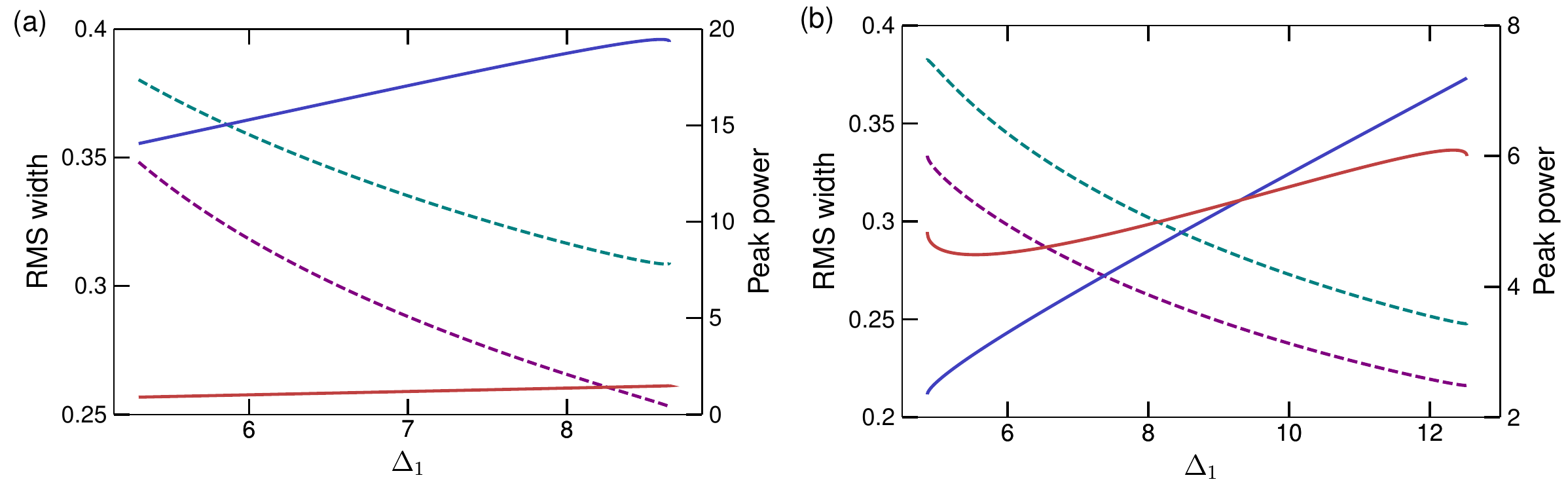}
    \caption{\textbf{Characterization of soliton width and peak power.} Individual variation of root-mean-square (RMS) width (dashed lines, left axis) and peak power (fully drawn lines, right axis) for the soliton components on the CS1 branch (a) and CS2 branch (b). Blue and green lines correspond to the FF while red and purple colors correspond to the TH. Only the stable part of the detuning interval is shown.}
    \label{fig:CS_nr}
  \end{minipage}
\end{figure*}
\begin{figure*}[!t]
\begin{minipage}{\linewidth}
    \centering
    \includegraphics[width=\linewidth]{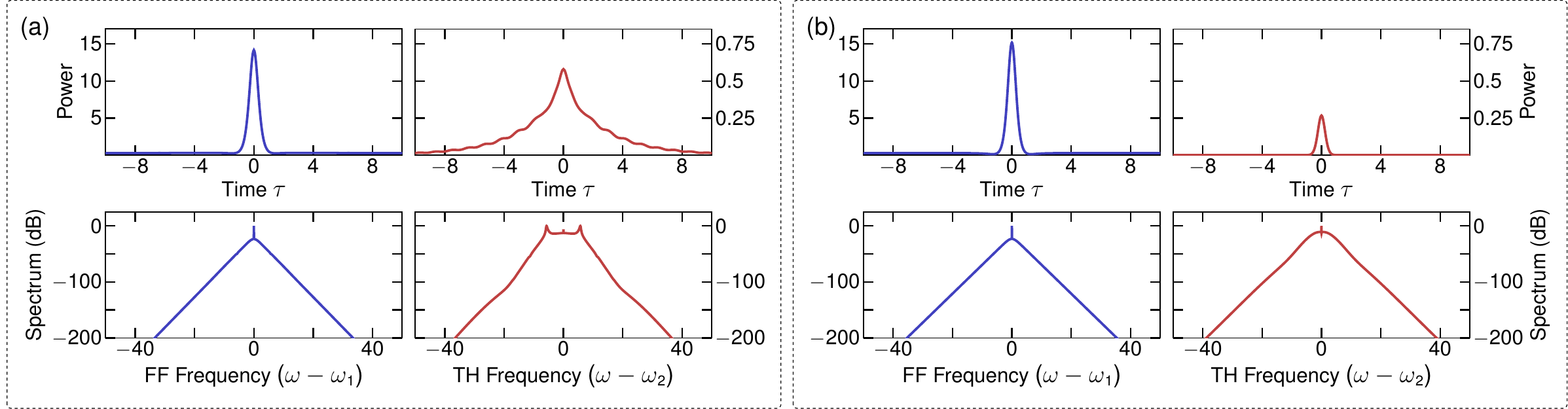}
    \caption{\textbf{Soliton profiles for mixed-dispersion conditions.} (a) Intracavity profiles (top) and spectra (bottom) of an isolated cavity soliton for $f = 3$, $\Delta_1 = 6$, $d = 0$, $\eta_1 = -1$ and normal TH dispersion $\eta_2 = 0.5$. (b) Isolated cavity soliton for the same normal dispersion case but with negative TH detuning $\Delta_2 = -3\Delta_1$. Blue and red colors denotes FF and TH components, respectively.}
    \label{fig:CS3n}
  \end{minipage}
\end{figure*}

To characterize the bifurcation structure undergone by the bistable solitons, we apply a path-continuation method based on a Newton-Raphson solver \cite{path_conti}. The results of these computations are depicted in Fig.~\ref{fig:CS_trace}. Figure~\ref{fig:CS_trace}(a) shows the modification of the FF energy of the CW, CS1 and CS2, captured through the $L_2$-norm $||A||^2=\int_{-t_R/2}^{t_R/2}|A(\tau)|^2d\tau$, as a function of the detuning $\Delta_1$ for $f = 3$. In Fig.~\ref{fig:CS_trace}(b) we show the same result after removing the homogeneous background field, which allows to better illustrate the organization of the solution branches. To do so we plot $||\partial_\tau A||^2$ against $\Delta_1$.  The bifurcation diagram associated with CS2 (in blue) originates from the saddle-node bifurcation of the CW solution SN$_h$. In contrast, the one corresponding to CS1 (in orange) forms an isola (i.e., a loop). Both curves exhibit a number of folding branches, with sets of both stable (fully drawn) and unstable (dashed) solutions. Some examples of temporal intensity profiles found along these curves are shown in the bottom row of Fig.~\ref{fig:CS_trace}, and correspond to the locations marked in Fig.~\ref{fig:CS_trace}(b). The stable soliton branches have turning points at the detuning values that correspond to the endpoints of the soliton steps, and the CS1 branch undergoes a Hopf-instability at the point Hopf$_{\rm CS1}$. The detuning range where CS1 and CS2 coexist (i.e., the soliton bistability region) is shown using a shadowed box.  

We further characterize the bistable solitons in Fig.~\ref{fig:CS_nr}, by plotting the variation of the root-mean-square (RMS) width of the temporal duration (dashed) and the peak power (fully drawn), of the soliton components for the fundamental and TH fields along the stable part of each branch. While we can observe a large difference in the peak power of the CS1 soliton components in Fig.~\ref{fig:CS_nr}(a), we see that their RMS time widths remain nearly the same. The higher TH power of the CS2 soliton in Fig.~\ref{fig:CS_nr}(b) makes it preferable as an operating state, in order to achieve optimal conversion efficiency for the TH comb. We observe similar trends of increasing peak power and simultaneously decreasing RMS duration, showing that the total power increases and the solitons become more energetic as the FF detuning grows larger. The soliton pair in panel (a) clearly has a smaller existence range, with the leftmost endpoint corresponding to the Hopf$_{\rm CS1}$ bifurcation, where the stable soliton transitions into a periodic breather state.

We emphasize that the observed bistability of dissipative CSs is due to the existence of two separate soliton attractors, and is different from the previously reported bistability mechanism of conservative spatial solitons in cavityless THG, that requires the presence of a phase-mismatch \cite{Sammut1998}. A similar bistability of vectorial dissipative solitons has recently been shown to occur for the physically distinct situation of nonlinear polarization mode coupling \cite{Averlant2017,Kostet2021}. Bistability of conservative solitons carrying the same power can also occur due to the appearance of two propagation constants when the nonlinear polarization has a certain functional dependence on the intensity, see Ref.~\cite{Kaplan1985}.

Admittedly, it may be challenging to find suitable nonlinear cavities where the simultaneous assumptions of zero walk-off and anomalous group-velocity dispersion for both FF and TH fields that were used to find the bistable CSs hold. Nevertheless, it has been previously demonstrated that both criteria can be satisfied under realistic experimental conditions. The temporal walk-off can, e.g., be made to vanish by considering different mode families, or sets of suitably chosen FF/TH frequencies on opposite sides of the zero-dispersion wavelength \cite{Hansson2018}; whereas anomalous TH dispersion may be obtained through dispersion engineering, or by exploiting avoided mode-crossings \cite{Savchenkov2012,Liu2014}. It is also likely that there may exist a similar bistability of dark dissipative solitons in the same parameter regime when the FF and TH fields both experience normal dispersion. However, the investigation of this case is beyond the scope of this work, and will be the subject of future studies.

Finally, we note that coupled dual solitary wave structures can also be found in the more common situation of normal TH group-velocity dispersion, $\eta_2 > 0$. Here the coupling is mainly perturbative, and we observe only short soliton steps that correspond to the upper branch of the homogeneous solution, see Fig.~\ref{fig:CS_sweep}(a). These solitons have a weak TH with a broad profile, as shown in Fig.~\ref{fig:CS3n}(a). They are similar to previously reported quadratic cavity solitons \cite{Hansson2018}, and may persist also in the presence of a large walk-off. The mixed-dispersion condition is clearly non-ideal, but it can be compensated by adjusting the TH detuning $\Delta_2$: specifically, supposing that $\Delta_2$ is not required to satisfy the condition of natural phase-matching ($\Delta_2 = 3\Delta_1$), but instead can be individually adjusted in order to change its sign. It is then possible to find coupled bright soliton pairs analogous to Fig.~\ref{fig:CS3}(a) also for normal TH dispersion, see for example the case of Fig.~\ref{fig:CS3n}(b).

\section*{Conclusions}

In conclusion, we have presented a theoretical model for optical Kerr frequency combs in a doubly-resonant and dispersive cavity system that is phase-matched for third-harmonic generation. We have reported conditions for achieving simultaneous dual-comb generation, and investigated a multistable regime that supports two types of bistable cavity solitons for anomalous dispersion. The parametric coupling between fundamental and third-harmonic waves allows the formation of simultaneous combs around multiple wavelengths, and is expected to be important for future applications of frequency combs in the visible and ultraviolet spectral range.

\section*{Methods}

\subsection*{Derivation of the mean-field model}

To derive the mean-field model, we start by expanding the intracavity fields in a power series as $A_m = A_m^{(0)} + \epsilon A_m^{(1)}$ and $B_m = B_m^{(0)} + \epsilon B_m^{(1)}$ where $\epsilon$ is a small parameter, c.f.~Ref.~\cite{Longhi1996}. The fields remain unchanged to the lowest order and the solution of Eqs.~(\ref{eq:cNLS1}-\ref{eq:cNLS2}) is $A_m^{(0)}(L) = A_m^{(0)}(0)$ and $B_m^{(0)}(L) = B_m^{(0)}(0)$. Inserting these solutions on the right hand side of the first-order equations and assuming that all terms are small, we can immediately carry out an integration to find that
\begin{align}
  A^{(1)}_m(L) = & A^{(1)}_m(0) + \left[-\frac{\alpha_{c1}L}{2} - i\frac{k_1''L}{2}\frac{\partial^2}{\partial\tau^2}\right]A^{(0)}_m + \nonumber\\
  & i\frac{\omega_1n_2(\omega_1)L}{c}\Big[Q_{13}\hat{\kappa}^* B^{(0)}_m({A^{(0)}_m}^*)^2 + \nonumber\\ 
  & \left(Q_{11}|A^{(0)}_m|^2 + 2Q_{12}|B^{(0)}_m|^2\right)A^{(0)}_m\Big], \\
  B^{(1)}_m(L) = & B^{(1)}_m(0) + \left[-\frac{\alpha_{c2}L}{2} - \Delta k'L\frac{\partial}{\partial\tau} - i\frac{k_2''L}{2}\frac{\partial^2}{\partial\tau^2}\right]B^{(0)}_m \nonumber\\ 
  & + i\frac{\omega_2n_2(\omega_2)L}{c}\Big[Q_{23}\frac{\hat{\kappa}}{3}(A^{(0)}_m)^3 + \nonumber\\
  & \left(2Q_{21}|A^{(0)}_m|^2 + Q_{22}|B^{(0)}_m|^2\right)B^{(0)}_m\Big],
\end{align}
where $\Delta k' = k_2'-k_1'$ is the group-velocity mismatch and $\hat{\kappa} = e^{i\Delta kL/2}\textrm{sinc}(\Delta kL/2)$. The boundary condition Eq.~(\ref{eq:BC1}-\ref{eq:BC2}) similarly become $A^{(0)}_{m+1}(0) = A^{(0)}_m(L)$ and $B^{(0)}_{m+1}(0) = B^{(0)}_m(L)$ to the lowest order, while the first order relations are given by
\begin{align}
  & A^{(1)}_{m+1}(0) = \sqrt{\theta_1}A_{in} - \left(\frac{\theta_1}{2}+i\delta_1\right)A^{(0)}_m + A^{(1)}_m(L), \\
  & B^{(1)}_{m+1}(0) = - \left(\frac{\theta_2}{2}+i\delta_2\right)B^{(0)}_m + B^{(1)}_m(L).
\end{align}
By combining the above expressions, and introducing a slow time variable $t$ we obtain a continuation of the map by setting $A_{m+1}(0) - A_m(0) \to t_R \partial A/\partial t$ and $B_{m+1}(0) - B_m(0) \to t_R \partial B/\partial t$ where $t_R$ is the roundtrip time. This leads to the following coupled system of mean-field evolution equations for the fundamental and third-harmonic fields
\begin{align}
  & t_R\frac{\partial A}{\partial t} = \left[-(\alpha_1+i\delta_1) - i\frac{k_1''L}{2}\frac{\partial^2}{\partial\tau^2}\right]A + i\frac{\omega_1n_2(\omega_1)L}{c}\times\nonumber\\
  & \left[Q_{13}\hat{\kappa}^*B(A^*)^2 + (Q_{11}|A|^2 + 2Q_{12}|B|^2)A\right] + \sqrt{\theta_1}A_{in}, \\
  & t_R\frac{\partial B}{\partial t} = \left[-(\alpha_2+i\delta_2) - \Delta k'L\frac{\partial}{\partial\tau} - i\frac{k_2''L}{2}\frac{\partial^2}{\partial\tau^2}\right]B + \nonumber\\ & i\frac{\omega_2n_2(\omega_2)L}{c}\left[Q_{23}\frac{\hat{\kappa}}{3} A^3 + (2Q_{21}|A|^2 + Q_{22}|B|^2)B\right],
\end{align}
where $\alpha_j = (\alpha_{cj}L+\theta_j)/2$ is the total roundtrip loss. We note that, besides the form of the nonlinearity, the resulting mean-field equations are analogous to models which have previously been obtained for doubly-resonant SHG frequency combs in $\chi^{(2)}$ resonators, see Ref.~\cite{Leo2016} for a derivation using an alternative approach.

The above equations are normalized with respect to the nonlinear coefficient and the time-scales for the losses and dispersion of the FF by rescaling $A \to \sqrt{\omega_1n_2(\omega_1)LQ_{11}/(c\alpha_1)}A$, $B \to \sqrt{\omega_1n_2(\omega_1)LQ_{11}/(c\alpha_1)}B$, $t \to (\alpha_1/t_R)t$ and $\tau \to \sqrt{2\alpha_1/(|k_1''|L)}\tau$. This choice of normalization results in the mean-field Eqs.~(\ref{eq:A}-\ref{eq:B}) of the main text. We note that the standard normalization for the LLE is obtained in the case that $B = 0$, which permits easy comparisons with previous Kerr comb results.

\subsection*{Eigenvalues of the characteristic equation}

The characteristic eigenvalue Eq.~(\ref{eq:M_matrix}) can be written in a factorized form as
\begin{equation}
  \left[(\lambda+1)^2+f_1\right]\left[(\lambda+\bar{\alpha})^2+f_2\right] = 3\rho p,
\end{equation}
where we have defined
\begin{align}
  & f_1 = q_1^2 - |p_1|^2 + 3\rho(|p_2|^2-|p_3|^2), \\
  & f_2 = q_2^2 - |p_6|^2 + 3\rho(|p_2|^2-|p_3|^2),
\end{align}
and
\begin{align}
  p = & \left(|p_2|^2-|p_3|^2\right)\left(q_1^2+q_2^2+(1-\bar{\alpha})^2-(|p_1|^2+|p_6|^2)\right)\nonumber\\
  & -\big[(p_1p_2^*+iq_1p_3)(p_2^*p_6^*+iq_2p_3^*)- \nonumber\\
  & (p_1p_3^*+iq_1p_2)(p_3^*p_6-iq_2p_2^*)+\textrm{c.c.}\big].
\end{align}
It can be shown that in the absence of coupling between the FF/TH fields the eigenvalues reduce to those of the LLE, i.e. $(\lambda+1)^2+f_1 = 0$. In the special case of equal losses and zero GVM ($\bar{\alpha} = 1$) the characteristic equation becomes biquadratic and has the explicit solution
\begin{equation}
  \lambda = -1 \pm \sqrt{-\frac{1}{2}\left(f_1+f_2\right) \pm \sqrt{3\rho p + \frac{1}{4}\left(f_1-f_2\right)^2}}.
\end{equation}
We note that in the absence of walk-off this solution becomes unstable for either of the three conditions $f_1+\alpha f_2+(1+\alpha)[\alpha+(1+\alpha)^2] < 0$, $\alpha(f_1-f_2)+[3\rho p+2\alpha(f_1+f_2)](1+\alpha)^2+\alpha(1+\alpha)^4 < 0$ and $(1+f_1)(\alpha^2+f_2)-3\rho p < 0$, c.f. Ref.~\cite{Hansson2018}.

\subsection*{Numerical methods}

The mean-field Eqs.~(\ref{eq:A}-\ref{eq:B}) have been numerically solved using a split-step Fourier method. The two equations are solved simultaneously by integrating them in a series of short alternating linear and nonlinear propagation steps with different basis. The dispersive linear step is performed in the frequency domain with $N = 2048$ modes and the forward and inverse transforms are implemented using Fast-Fourier Transforms (FFTs). The nonlinear step is meanwhile performed in the time-domain using a 4th-order Runge-Kutta algorithm. \\\\
The soliton bifurcation diagrams shown in Fig.~\ref{fig:CS_trace} have been obtained through a path-continuation algorithm, by computing the stationary solutions of Eqs.~(\ref{eq:A}-\ref{eq:B}) (by setting $\partial_tA=\partial_tB=0$) and varying the detuning $\Delta_1$. To do so we have recast the stationary version of the equations into the eight-dimensional dynamical system
 \begin{equation}
 	\frac{d\bar{u}}{d\tau} = \bar{F}(\bar{u}), \label{eq:Fu}
 \end{equation}
with $u_1(\tau)=U(\tau)={\rm Re}[A]$, $u_2(\tau)=V(\tau)={\rm Im}[A]$,  $u_3(\tau)=W(\tau)={\rm Re}[B]$,  $u_4(\tau)=Z(\tau)={\rm Im}[B]$,   $u_5(\tau)=U'(\tau)$,  $u_6(\tau)=V'(\tau)$, $u_7(\tau)=W'(\tau)$, $u_8(\tau)=Z'(\tau)$,
\begin{align}
		& F_m = u_{m+4}, \qquad\qquad m=1,\ldots, 4 \nonumber\\
		& F_5 = \eta_1^{-1}\left[-u_2-\Delta_1 u_1+ \mathcal{N}_2(u_1,u_2,u_3,u_4)\right], \nonumber\\
		& F_6 = \eta_1^{-1}\left[u_1-\Delta_1 u_2-\mathcal{N}_1(u_1,u_2,u_3,u_4)+f\right], \nonumber\\
		& F_7 = \eta_2^{-1}\left[-\alpha u_4-\Delta_2 u_3+\mathcal{N}_4(u_1,u_2,u_3,u_4)\right], \nonumber\\
		& F_8 = \eta_2^{-1}\left[\alpha u_3-\Delta_2 u_4-\mathcal{N}_3(u_1,u_2,u_3,u_4)\right],
\end{align}
and
\begin{subequations}
\begin{align}
    \mathcal{N}_1(U,V,W,Z) &= \kappa\left[2UVW-Z(V^2-U^2)\right] \nonumber\\
    & -V\left[U^2+V^2+2\sigma(W^2+Z^2)\right],
\end{align}
\begin{align}
    \mathcal{N}_2(U,V,W,Z) &= \kappa\left[2UVZ+W(U^2-V^2)\right] \nonumber\\
    & +U\left[U^2+V^2+2\sigma(W^2+Z^2)\right],	
\end{align}
\begin{align}
    \mathcal{N}_3(U,V,W,Z) &= -\rho\kappa(3U^2V-V^3) \nonumber\\
    & -3\rho Z\left[2\sigma(U^2+V^2)+\mu(W^2+Z^2)\right],
\end{align}
\begin{align}
    \mathcal{N}_4(U,V,W,Z) &= \rho\kappa(U^3-3V^2U) \nonumber\\
    & +3\rho W\left[2\sigma(U^2+V^2)+\mu(W^2+Z^2)\right].
\end{align}
\end{subequations}
This allows us to compute soliton states as a boundary value problem, imposing Neumann boundary conditions at $0$ and $t_R/2$, and utilizing the open distribution software package AUTO-07p \cite{AUTO}.
The linear stability of these states is ascertained through computation of the eigenvalues of the Jacobian matrix associated with the system (\ref{eq:Fu}). 

\section*{Data availability}

The data that support the findings of this study are available from the corresponding author upon reasonable request.

\section*{Code availability}

The numerical codes used for this study are available from the corresponding author upon reasonable request.

\section*{Acknowledgements}

This work has received funding from the European Union’s Horizon 2020 research and innovation programme under the Marie Skłodowska-Curie projects MOCCA (814147) and NOSTER (101023717), and the Italian Ministry of University and Research project WASHING (R18SPB8227).

\section*{Author contributions}

T.H. developed the theoretical model and wrote the manuscript. P.P.R. performed the soliton bifurcation analysis and S.W. supervised the work. All authors discussed and contributed to the theoretical interpretation of the results.

\section*{Competing interests}

The authors declare no competing interests.

\end{document}